\documentclass[12pt]{article}
\usepackage{subfigure}
\usepackage{epsfig}
\begin{document}
%
\title{Signatures of Precocious Unification in Orbiting Detectors}
\author{G.~Domokos and S.~Kovesi-Domokos\\
Department of Physics and Astronomy\\
The Johns Hopkins University\\
Baltimore, MD 21218\thanks{e-mail: skd@jhu.edu}\\
and\\
William~S.~Burgett and Jason Wrinkle\\
Department of Physics\\ University of Texas at Dallas\\
Richardson, TX 75083\thanks{e-mail: burgett@utdallas.edu}}
\date{Revised version.}
\maketitle
\abstract{It has been conjectured that the string and unification scales 
may be substantially lower than previously believed, perhaps a few TeV.
In scenarios of this type, orbiting detectors such as  OWL or AIRWATCH can 
observe
spectacular phenomena at trans-GZK 
energies. We explore measurable signatures of the hypothesis that
trans-GZK air showers (``anomalous showers'') are originated by strongly 
interacting neutrinos.
The results of a MC simulation of such air showers are described. A distinction
between proton induced and ``anomalous'' showers is possible once
a substantial sample of trans-GZK showers becomes available.
\nopagebreak \begin{flushleft}
 \end{flushleft}}
\maketitle
\section{Introduction}
The existence of trans-GZK cosmic rays is now reasonably well established:
various detectors such as Fly's Eye, Agasa, and HiRes have collected 
over 20 events with primary energy exceeding $10^{20}$eV. Based on
present flux estimates, new experiments such as OWL, AIRWATCH,  
and the Pierre Auger observatory 
are expected to collect a total of over 
$10^3$ events/year in this energy range.

The existence of trans-GZK events is a major puzzle since no source of 
such energetic particles has been plausibly identified within a distance of about
50~Mpc from the solar system. A brief summary of attempts to identify
sources of trans-GZK events can be found in a recent review of Sigl\cite{sigl}.

In previous work we have conjectured that neutrinos of energy 
$E_{0}\sim10^{20}$ eV may acquire a strong interaction at the CM energies of 
the collision with an air nucleus while penetrating the CMBR essentially 
uninhibited~\cite{domonuss, skd, prl}. 
Here, we summarize the essential features of this scenario.
Consistency of string models requires that the strings ``live'' in a spacetime
of dimension higher than the observed macroscopic spacetime (typically, $d\ge10$).
It has been observed that in such a situation the characteristic string scale
$M_s$ and the ({\em macroscopic}) Planck scale need not be the same, 
see~\cite{lykken, dimo, dienes}\footnote{Due to to the large body of
literature on the subject, we cite only the first works of the 
relevant authors on this topic.}.
In fact, the string scale can be considerably lower than the macroscopic
Planck scale, perhaps of the order of 10 TeV or so. The exact relationship
between the fundamental string scale and the macroscopic Planck scale depends
on how one hides the extra dimensions. For example, one can think of compactified
extra dimensions (as in ref.~\cite{lykken, dimo}) or of a Randall-Sundrum
mechanism~\cite{rs1}.

The transition from  the Standard Model regime to the regime
governed by string physics  is likely to be fast
due to the rapidly rising level density in a string model~\cite{prl}.
There is a large degree of uncertainty in the details of such models.
In particular, the characteristic energy scale (the string scale) can lie
anywhere between a few TeV to perhaps a few hundred TeV. Hence, an analysis of
experimental/observational data exploring these ideas has to satisfy 
the following criteria:
\begin{itemize}
\item It should concentrate on identifying 
the most robust features of models incorporating new (``stringy'') physics;
details vary from model to model, and it is difficult to foresee how a future
more complete theory will appear. Nevertheless, abstracting the important
features of presently existing models provides guidance for future observations.
\item Due to the uncertainty in establishing the characteristic
energy scale, and recalling that the CM energy is in the range of a few hundred TeV for
the trans-GZK events, it is necessary to investigate ultra-high energy cosmic ray (UHECR) 
data in addition to data from accelerator experiments. In developing new experiments
for collecting UHECR data, the currently preferred detection system is an orbiting 
detector such as OWL or Airwatch because of the large target area available for observation 
as well as the capability to track the complete evolution of an atmospheric shower.
\end{itemize}

The approach adopted for this study was based on the following two main assumptions.
\begin{itemize}
\item The string and unification scales are close to each other
or are equal.  We denote the characteristic energy scale governing the onset of the
string regime by $\sqrt{S}$. Due to uncertainties in model building at present, 
we set $S\approx M_{s}^{2}\approx 1/\alpha^{'}$, thus ignoring a string coupling constant
of order 1, and where $1/\alpha^{'}$ is the inverse Regge slope of the string model
related to the string tension in the usual manner.
\item Due to the excitation of the string degrees of freedom, there is a rapid
transition between the regime where physics is described by the Standard Model
and the  string regime with unified interaction strength.
\end{itemize}
A detailed investigation of the sensitivity of MC simulations to particle
physics uncertainties was carried out by Mikulski~\cite{pault}. He found that
for the case discussed here, the shower profiles and the fluctuation pattern
are insensitive to the precise form of the transition between 
the Standard Model and string regimes, as well as to the precise form of 
the level density in the transition regimes, as long as the latter was a
rapidly (typically, exponentially) rising function of  $\sqrt{s}$.

In what follows, we provide an overview of the ALPS Monte Carlo 
simulation and the theoretical modeling used to obtain the results. The last
section summarizes the key points of the work with an emphasis on the relevance
to future orbiting detector experiments.

\section{The simulation and its results}

\subsection{The ALPS simulation}
A detailed description of the ALPS (Adaptive Longitudinal Profile 
Simulation) Monte Carlo
program created by Mikulski is contained in ref.~\cite{pault}. ALPS 
simulates the longitudinal development of a shower, and this is adequate for
the high energies used in this study.  A particularly useful feature of ALPS  
is its relatively fast execution time to generate air shower histories. 
It is easily modified to 
accept user-defined input physics beyond the Standard Model, and is 
adaptable as a front end generator to detector Monte Carlo simulations. 
ALPS is already being used by members of the OWL collaboration for 
cosmic ray signal and detector characterization.

More specifically, the starting point for the hadronic cascades 
is Approximation A, {\em i.e.}, the interaction
length is independent of energy, and the inclusive cross section  is
dependent only on the ratio of outgoing particle energy
to incident particle energy (Feynman scaling). In general, particles created in an air shower
are simulated and tracked until their energy falls below a user-defined threshold
(set to $1/1000$ of the incident primary energy for this study)
after which subshower parameterizations are introduced.
The distribution of particles produced
in the cascade below the above mentioned threshold is  represented by a modified Gaisser-Hillas
distribution.  Corrections due to nuclear target effects and scaling violations
are introduced both in the simulation and in the parameterized subshowers. 
The fitting parameters in the profile are dynamically adjusted by the
program to satisfy goodness-of-fit criteria based on tracking selected 
subshowers. The electromagnetic cascades are implemented using a modified Greisen
parameterization algorithm. Finally, there is a correction
for the reduction of the bremsstrahlung and pair creation cross sections due to
multiple scattering in the atmosphere (the Landau-Pomeranchuk-Migdal or LPM effect).

\subsection{Theoretical considerations and modeling}
For the purpose of exploring the conjectured ``new physics'' described in
ref.~\cite{prl}, it was assumed that as long as $\sqrt{s}\geq \sqrt{S} \approx M_{s}$,
an interaction produces an equal number of leptons and quarks. Once the 
primary energy falls below   the string scale, $\sqrt{S}$,
the shower evolves according to
Standard Model physics. It was found that a step function-like
onset of the precociously unified physics gives a description indistinguishable
from an exponentially rising level density which for $\sqrt{s} \gg M_{s}$
levels off due to unitarity corrections\footnote{Strictly speaking, a
$\theta$-function step in the cross section violates unitarity, since the real part
of the forward elastic amplitude develops a logarithmic singularity.
However, any smoothing of the $\theta$-function removes the singularity
and unitarity can be restored. There is no harm done if a step function
is used in the total cross section: the latter  depends on the imaginary part of 
the amplitude only.}.
Hadronization of the quarks both in the
string and Standard Model regimes is assumed to be consistent with a
conventional splitting algorithm.
Due to uncertainties in the models, at present no predictions can be made 
about how fast the various coupling constants are running with energy, but it is 
expected that they change faster than in the Standard Model~\cite{dienes}. 
The exact functional forms of the $\beta $ functions in the renormalization
group (RNG) equations that determine the energy scaling of the coupling constants
depends on the spectrum of Kaluza-Klein excitations and on the specific string theory. 
For this reason, we extrapolated the strong cross sections into the
energy range of $\sqrt{s} \approx 500$TeV using the fit of 
Block~{\em et al.}~\cite{block},
and set the cross section in the string regime to half of that value, due 
to uncertainties in the extrapolation. We prefer to err on the conservative 
side.
Again, the results do not depend critically on the precise magnitude of the
cross section in the string regime.

Relative multiplicities follow from the above assumption that
any particle produced has a $50\%$ probability of being a lepton or quark as long as
the energy is larger than $\sqrt{S}$. Once energies 
fall below the characteristic scale,  lepton interaction produces a multiplicity of 2 (counting
leptons and photons on an equal footing), whereas a quark, after hadronization, produces high
multiplicities in each interaction. At the energies considered
here, a muon emits a bremsstrahlung
photon almost at the same rate as an electron. Consequently, 
 there is an excess of the  low multiplicity component in the ``anomalous''
 shower ({\em i.e.},
generated by a neutrino with precociously unified
interactions). This contrasts 
with proton induced showers where the source of the leptonic component is the decay,
$\pi^{0}\rightarrow \gamma \gamma$. At the present level of accuracy, photoproduction of
pions and other mesons can be neglected. 

\subsection{Average shower properties}
The average profiles used for this study were derived such that shower development could
be characterized independently of the relative position with respect to the detector. 
Thus, the impact parameter $b$ is selected to parameterize the geometry of the shower.
Recalling that $b$ is  the distance of closest approach of the shower axis 
measured from the center of the Earth,
we present results only for $b > R_{\oplus}$. Then, writing $b = R_{\oplus} + h$,
the altitude $h$ is the distance of closest approach measured from the
surface of the Earth. Horizontal distances parameterizing the longitudinal shower
development are defined to be measured along the shower axis, 
taking the point of closest approach as zero distance. Since all 
relevant showers take place in the upper atmosphere, an exponential atmosphere
was used with a scale height of $h_{0}=6.4$ km yielding an
adequate approximation to Shibata's parameterization~\cite{gaisser}. 

\begin{figure}[tb]
\centering
\epsfig{figure=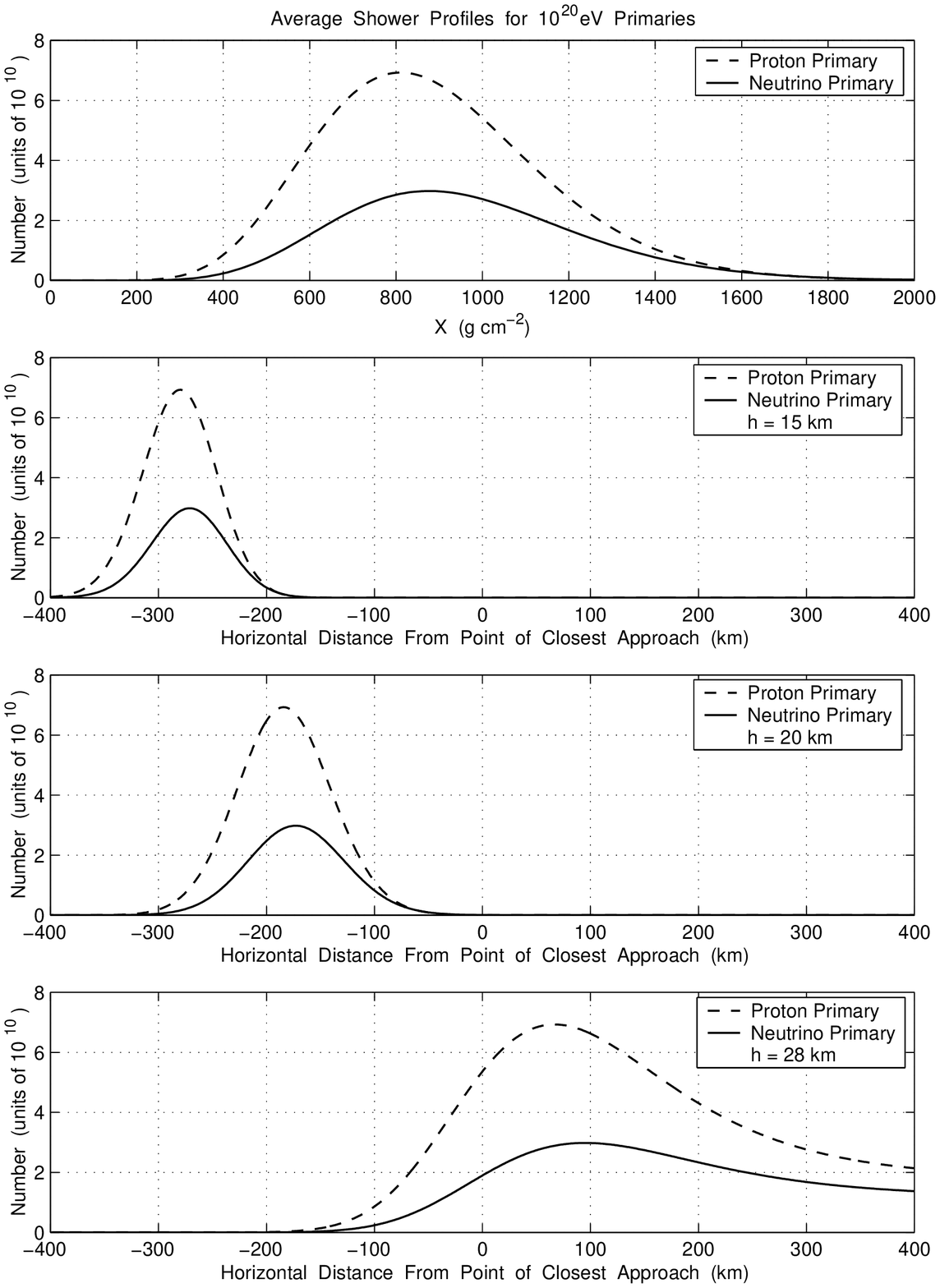, width=0.9\textwidth}
\caption{Comparison of average shower profiles for proton and neutrino 
induced showers as a function of column density and as a function of height
above the Earth for a neutrino interaction cross section equal to half
the proton value.}
\label{averageprofile}
\end{figure}

\begin{figure}[tb]
\centering
\epsfig{figure=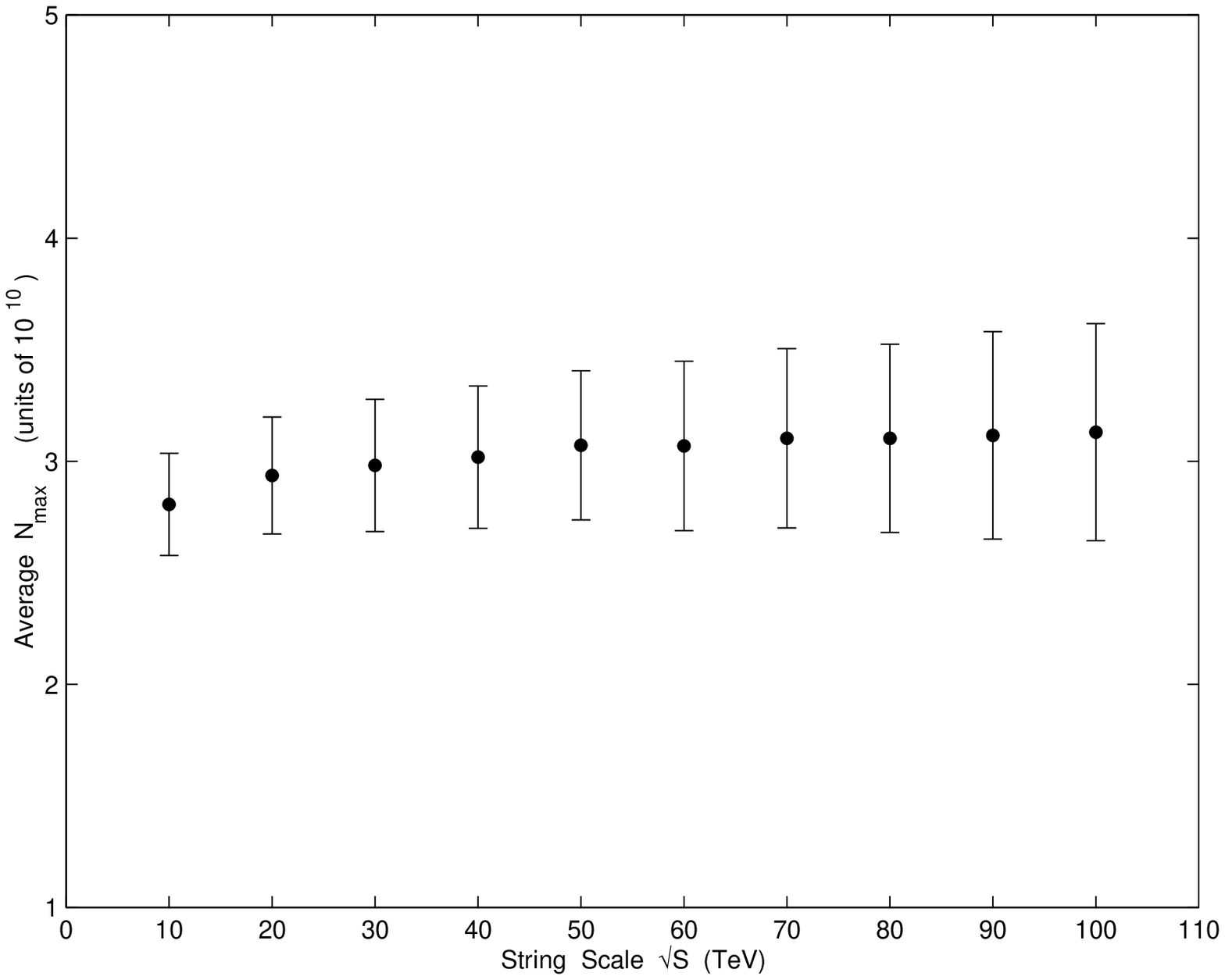,width=0.9\textwidth}
\caption{The number of electrons at shower maximum as a function of the
string-motivated precocious unification scale $\sqrt{S}$ for a neutrino
interaction cross section equal to half the proton value
for sets of 1000 Monte Carlo runs.}
\label{nmaxstringscale}
\end{figure}

The average shower profiles shown in Figure~\ref{averageprofile} are for $E_{0}= 10^{20}$ eV 
proton and neutrino primaries incident on air nuclei
(corresponding to $\sqrt{s}\approx 400$ TeV). The neutrino primaries are then
``strongly interacting neutrinos'' for precocious 
unification thresholds below $\sqrt{S} \sim 400$ TeV.  As an example, the
neutrino induced anomalous showers shown here were generated for 
$\sqrt{S}=30$ TeV and $\sigma_{\nu} = \sigma_{p}/2$. Figure~\ref{averageprofile}(a) shows 
the two profiles
as a function of column density while Figures~\ref{averageprofile}(b-d) present the profiles
as a function of altitude above sea level to explicitly exhibit atmospheric
effects. There is considerable broadening in the shower development at higher 
altitudes due to the low density of air as is evident in the profiles at 
$h = 28$ km. As shown in Figure~\ref{nmaxstringscale}, the number of electrons at the shower 
maximum, $N_{max} \approx 3\times 10^{10}$, is  only weakly dependent on the 
precise value of the string scale. 


\subsection{Fluctuations in shower development}
The average shower profiles displayed in Figure~\ref{averageprofile} for neutrino induced
showers reveal an interesting feature when compared with the proton induced events.
The multiplicity around $X_{max}$ is about half of the value for the
proton induced shower due to the unified forces allowing a substantial
portion of the primary energy to be channelled into prompt lepton production with 
the lepton interaction cross sections and multiplicities being smaller than 
those for hadronic channels. Consistent with this picture is the result that the
electron deficiency {\em increases} with {\em decreasing} $\sqrt{S}$: for a lower
characteristic energy scale, the prompt lepton production due to unification
occurs over a longer interval of the shower development after the first interaction.

Although the average profile of a neutrino induced shower is different
from a Standard Model proton induced shower, it is hard to distinguish between the two types
on an event-by-event basis. While it is true that {\em on average} $N_{\rm{max}}$ in
a proton induced shower is roughly 2-3 times as large as for a neutrino induced event, and
the development of the neutrino induced showers is somewhat slower compared to those induced
by protons, fluctuations are likely to smear out such differences in any given shower.

However, a statistical analysis of a sample of showers should reveal a significant difference
between neutrino and proton induced showers. Qualitatively, an important shower property  
is that smaller cross sections and smaller average multiplicities in individual interactions
lead to larger fluctuations in the shower development. This qualitative expectation 
is borne out by the simulation results. 
\begin{figure}[tb]
\centering
\epsfig{figure=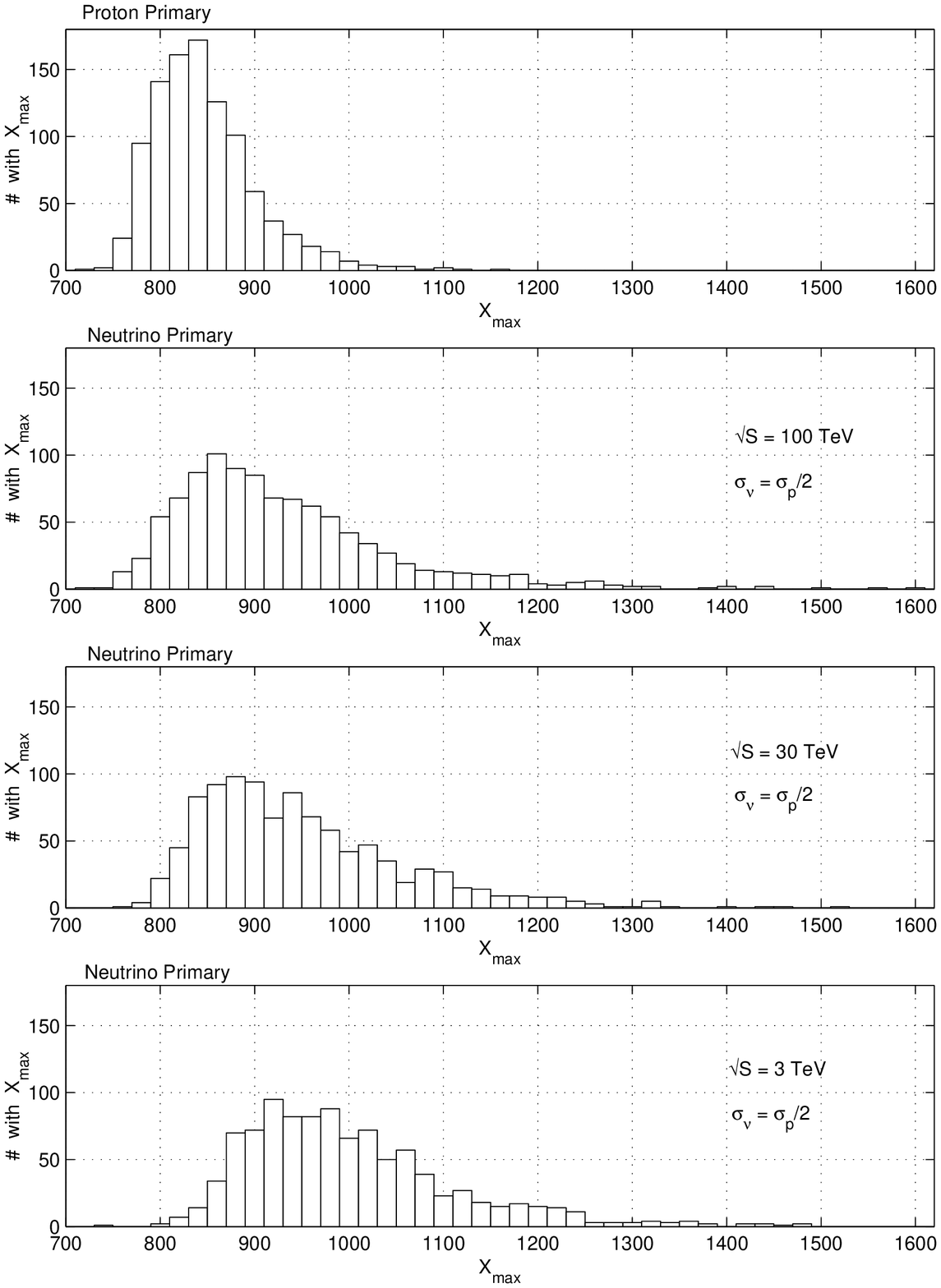,width=0.9\textwidth}
\caption{Distribution of shower maxima as a function of column density
for precociously unified $10^{20}$ eV neutrino induced showers at 3 string scales compared
to a conventional Standard Model $10^{20}$ eV proton induced shower.}
\label{xmaxdistribution}
\end{figure}
Figure~\ref{xmaxdistribution} displays the distribution of
the position of the shower maximum, $X_{max}$,
for various values of the characteristic scale, $\sqrt{S}$. For comparison, the same
distribution is shown for normal showers generated by protons in the absence of new
physics. Clearly, the width of the distribution in $X_{max}$ for all the showers 
containing precociously unified interactions
is considerably larger than those induced by protons.
This is more evident if one plots the second central moments of the distributions versus the
mean $X_{max}$ as in Figure~\ref{standarddeviation}:
there are clearly two distinct regions
for proton and neutrino induced showers in this parameter space.
One also observes that
given sufficient statistics, the distribution in $X_{max}$ gives a hint about the
magnitude of $\sqrt{S}$: lower characteristic energies give rise to longer 
tails in the distribution.

Preliminary calculations also indicate that variations of the cross section at
unification do not greatly affect the qualitative features of the results presented
here. For example, if the cross section at unification is assumed to be equal to
the extrapolated hadronic value used in this paper, $\langle X_{max} \rangle$ gets
somewhat closer to the value for proton induced showers, and the rms fluctuations
about $X_{max}$ also decrease. However, the shower does not become statistically
equivalent to the proton case. The difference arises because after the first few
interactions in the neutrino showers, aproximately half the energy is distributed
among leptons, and when the energy drops below the string threshold, the leptons
contribute to shower development through lower multiplicity interactions
compared to the hadronic channels.
\begin{figure}[tb]
\centering
\epsfig{figure=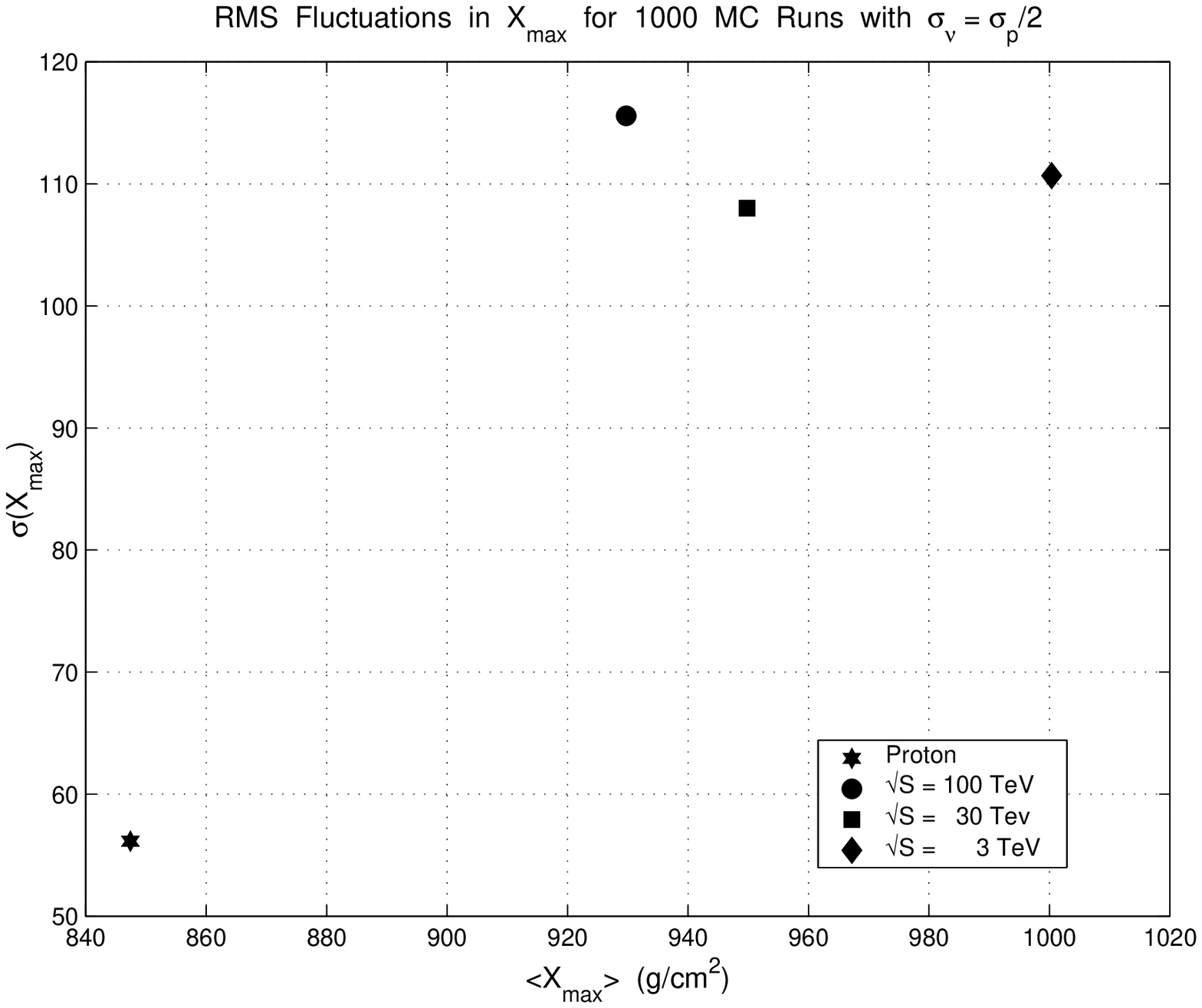,width=0.9\textwidth}
\caption{Standard deviations of the $X_{max}$ distributions shown in Figure\ref{xmaxdistribution}
for proton and neutrino induced 
showers at various values of $\sqrt{S}$ and for a primary energy $E_{0} = 10^{20}$ eV.}
\label{standarddeviation}
\end{figure}
\section{Discussion}
We find that orbiting detectors (and to some extent, any detector of the Fly's Eye type)
can yield information about the presence or absence of certain types of 
new physics in trans-GZK showers. Two features are important from this point of view:
\begin{itemize}
\item The detector has to collect a substantial sample of trans-GZK events, since
the difference between proton and neutrino induced showers is not sufficiently
large for a distinction on an event-by event basis. This is primarily due to the fact that
at any reasonable value of $\sqrt{S}$, the physics of the shower eventually becomes dominated by ``low
energy'' Standard Model phenomena, and, hence, the largest number of particles is  generated 
in the latter regime of shower development. A particular advantage of orbiting detectors
is the capability for observing showers at a broad range of impact parameters,
thus allowing an estimate of the magnitude of the cross sections.
\item It appears that the shower development is rather insensitive to finer details of
the models discussed in this paper. 
This is actually  fortunate given the theoretical uncertainties at the present 
stage of model building. It is not yet clear how the 
analysis of data can be refined until  more detailed and experimentally constrained models
become available.
However, the analysis proposed here is  apparently robust. Once a sample size of
greater than $\sim$100 events with well-determined profiles becomes available,
a statistical analysis will reveal the presence or absence of the new physics proposed
here.
\begin{figure}
\centering
\epsfig{figure=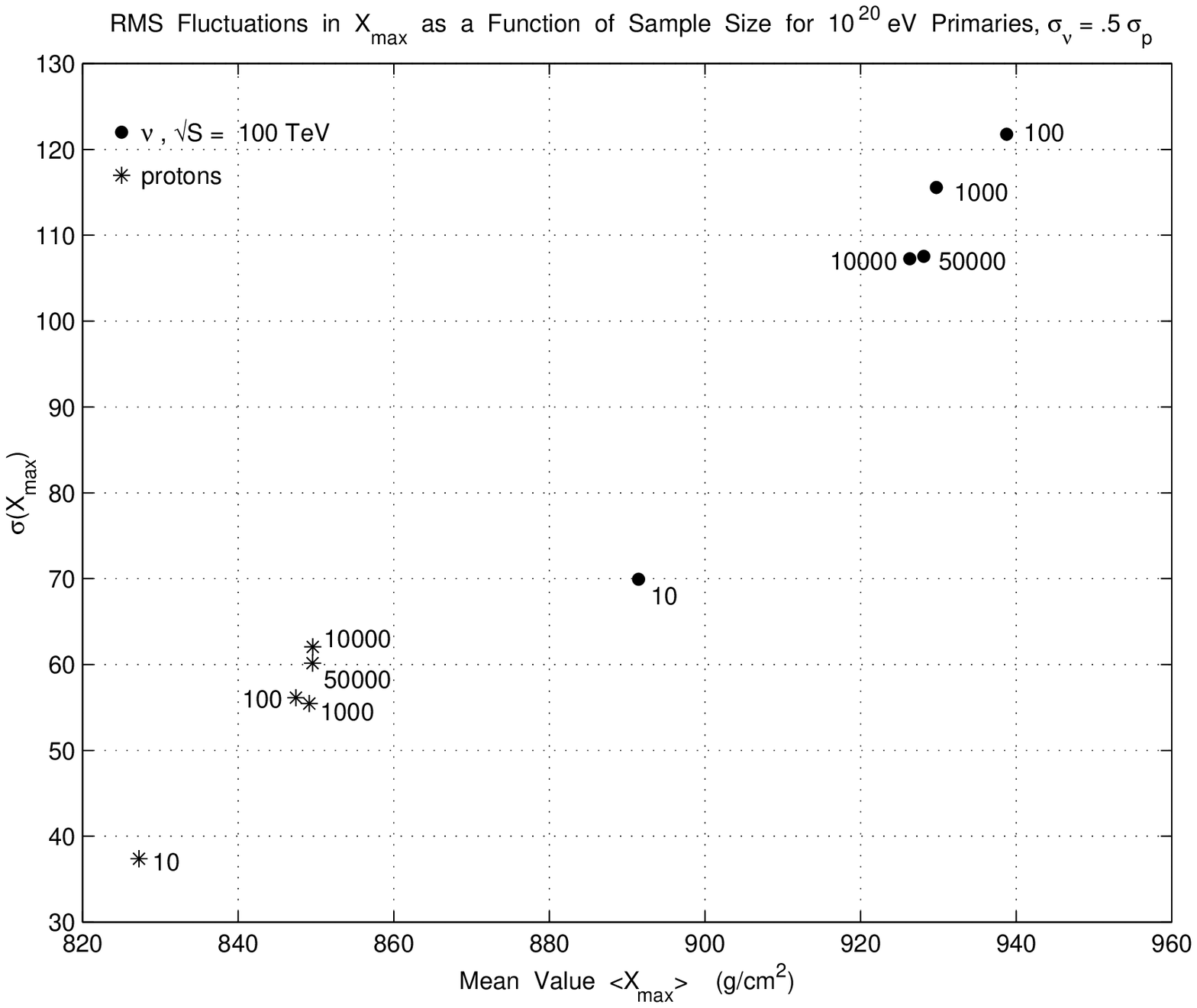,width=0.9\textwidth}
\caption{The measured standard deviations plotted against the average $X_{max}$
for various sample sizes (the number adjacent to a data point indicates the number
of MC runs over which averages were computed). Note that after only
$\sim100$ events, the average $X_{max}$ 
and the standard deviation $\sigma(X_{max})$ converge to $\approx 1\%$ and $10\%$
of their final values, respectively.}
\label{samplesize}
\end{figure}
This is illustrated in Figure~\ref{samplesize} where an event size of about
100 already gives a clear distinction between proton and neutrino induced showers.
The result does not vary dramatically with
a further increase of the number of events measured. 
Although the mean $X_{max}$ values for the neutrino events change by only about $1 \%$
between the 100 and 50,000 run sample sizes, the standard deviations, as expected,
settle more slowly changing by about 10$\%$ from 100 to 50,000 runs. Thus,  
very large sample sizes are required to bracket final values of the standard deviations 
to better than $1 \%$.
\end{itemize}

\vspace{1cm}
\large{\noindent \textbf{Acknowledgements}}\\*[0.3cm]
\small
We thank John Krizmanic for several useful discussions regarding the design of OWL. In
particular, our choice of the significant impact parameters was based on his estimate
of the intensity of air fluorescence as a function of the height above the surface of the
Earth. We are also indebted to Paul Mikulski for sharing his insight into the ALPS
program he authored.

\end{document}